\documentclass[apj, revtex4]{emulateapj}
\usepackage{amsmath}
\usepackage[T1]{fontenc}
\usepackage{float}
\usepackage{booktabs}
\usepackage{longtable}
\usepackage{epsfig,natbib}
\usepackage{enumerate}
\usepackage{multirow}
\usepackage{subfigure}
\usepackage{gensymb}
\usepackage{morefloats}
\usepackage{lscape}
\usepackage{graphicx}
\usepackage{color}

\begin{document}

\slugcomment{Draft Version; not for circulation}
\title{Refutation of K2-256b and Rejection of 130 Unconfirmed Transiting Planet Candidates with Gaia DR3 NSS}

\author{Thomas W. Tarrants\altaffilmark{1}}
\author{Elvis O. Mendes\altaffilmark{2}}
\altaffiltext{1}{Corresponding Email: twtarrants@gmail.com}
\altaffiltext{2}{Corresponding Email: elvisoliveiramendes@gmail.com}

\begin{abstract}
Much of the exoplanet discovery efforts over the next several years are largely tasked with finding candidates for the upcoming Ariel mission. This is a role that TESS is well-suited for. Radial velocity follow-up is needed to confirm the planetary nature of these systems, as many of its planet candidates turn out to be eclipsing binary systems with small stellar secondaries. Focused Doppler follow-up to obtain these radial velocities is expensive. The Gaia mission's radial velocity measurements on its target stars are not adequately precise for a general search for orbiting planetary companions. While the RV data has yielded thousands of spectroscopic binary systems, only an extreme minority of these reach into the planetary-mass regime. Nonetheless, they do detect eclipsing binary systems that can masquerade as transiting hot Jupiters. In this work, we compare the Gaia DR3 Non-Single Stars catalogue to the current planet candidate list from TESS and determine several of these candidate planetary systems are actually eclipsing binaries. We find $\sim130$ eclipsing binaries among the TOI and EPIC lists that do not appear to be in the literature, including the previously statistically validated planet K2-256 b. This work illustrates the usefulness of \textit{Gaia} to exoplanet validation efforts and will help guide follow-up efforts for discovering new transiting planets for the Ariel mission by avoiding misdirecting valuable telescope time.

\end{abstract}


\section{Introduction}
\label{I}

With the advent of space-based transit surveys like \textit{Kepler} and \textit{TESS}, the floodgates have been opened: thousands of candidate transiting planets have been identified, allowing for extrasolar planets to be understood at the \textit{population} level. Nevertheless, important to maintaining a high-fidelity sample of extrasolar planets is excluding astrophysical false-positives -- eclipsing binary stars that produce transits that appear to mimic those of extrasolar giant planets.

Many more candidate planets have been identified than can pursued with follow-up ground-based Doppler detection, and many of the smaller planet candidates identified by these missions are inaccessible to Doppler detection without a lengthy and intensive dedication of telescope time. There is simply not enough high-precision spectrometers to go around.

This problem is compounded by non-negligible false-positive rates. \cite{Santerne2012} used \textit{SOPHIE} radial velocity data to estimate a false positive rate of $34.8\pm6.5$ percent for hot Jupiters from \textit{Kepler}. Concerns over the need to identify genuine planets from false positives encouraged the development of statistical validation techniques where false positive scenarios were ruled out to sufficiently high levels of confidence that a candidate planet without a Doppler detection could be argued to be "validated" with false positive probabilities below a sought threshold \citep{Morton2012}, typically on the order of $10^{-3}$.

Celestia\footnote{\\Website: https://celestiaproject.space/ \\ Discord: https://discord.gg/gYagY6W2Gv} is a free, flexible, editable 3D Space Simulator that allows users to view solar and extrasolar planetary objects, moons, comets, etc. There is a community of content creators for it that add extrasolar planets as they're discovered, various eclipsing binaries, and other phenomena of interest. It was noticed by this community that some of the new TESS planet candidates were overlapping with known eclipsing binary stars, especially from the EBLM project \citep{Triaud2017}. This motivated an attempt to identify more systems out of a desire to help exclude candidate planets that were clearly astrophysical false positives.

The Gaia DR3 NSS SB1 catalogue \citep{GaiaDR3NSSSB1} contains over a hundred thousand spectroscopic binaries, many of which are no-doubt transiting. Combining this catalogue with the \textit{TESS} list of transiting planet candidates regularly updated at the Exoplanet Follow-up Observing Project (ExoFOP) website provides a way of constraining the nature of these candidate planets and in many cases determining whether if they can be attributed to astrophysical sources, such as eclipsing binaries. For thoroughness, we also performed the same check on the list of planets and planet candidates identified by the \textit{Kepler-K2} mission. We found over 130 matches between Gaia SB1 systems and planet candidates where the Gaia system could be positively matched to the source of the transits identified by these two missions, including most interestingly a planet that had been statistically validated at K2-256.

\section{Methodology}

Two .csv files, one containing the \textit{TESS} Object of Interest (TOI) list which was downloaded from the ExoFOP, and the other containing a list of \textit{Kepler-K2} candidates from the NASA Exoplanet Archive. The Gaia DR3 NSS SB1 list was then downloaded from VizieR. The \textit{TESS} pixel size is 21 arcseconds, but given the potential for contamination from nearby background eclipsing binaries contributing flux within the \textit{TESS} PSF of a given candidate transiting planet system, we searched for Gaia NSS SB1 targets within $\sim$5 times this radius limit (250 arcseconds). A search box around the \textit{TESS} target was defined by $\alpha_\text{G} - (\alpha_\text{T} / \cos{\delta_\text{T}}) < 0.1$ and $\delta_\text{G} - \delta_\text{T} < 0.1$, where the subscripts G and T denote the coordinates for the \textit{Gaia} and \textit{TESS} systems, respectively and the $\cos{\delta_\text{T}}$ accommodates the geometric distortion that arises from a two-dimensional coordinate system projected onto a spherical sky, resulting in "pinching" toward the celestial poles. We used the same radius for the search for matches for \textit{Kepler-K2} systems for consistency. Any \textit{Gaia} systems within this box were taken as a candidate match to the \textit{TESS} system. This search therefore includes many systems that do not have the same \textit{Gaia} DR3 identifier as the TIC system presented, allowing us to identify nearby eclipsing binary blends.

\subsection{Candidates with Known Periods}

We accept as a match a Gaia SB1 solution whose period is within a factor of $10^{-3}$ of the sought period (or a factor of two of the sought period, as eclipsing binaries can often appear to have a photometric period half the radial velocity period), or if the Gaia DR3 source ID matches that of the star with the transiting planet candidate. If a match is found, there are two ways forward:

a) The Gaia DR3 NSS SB1 period matches the period of the transiting planet candidate. In this case, we can safely ascertain that the system detected in transit is an eclipsing binary that has produced the radial velocity variations that led to its detection by Gaia.

b) The Gaia DR3 NSS SB1 period does not match the period of the transiting planet candidate. In this case, multiple possibilities present themselves.

It could be that the system is, for example, an eclipsing binary with approximately equal-brightness stellar components. For circular orbits this can produce a transit period that is half the true period, and consequently half the period reported with radial velocity. For eccentric orbits, it's possible that an eclipse and a transit occur in the same observation window with a ratio only related to the true period via the orbital eccentricity and longitude of periapsis. In this case, the true period and the reported period from transits could have a non-trivial ratio.

One conceivable scenario is that an unresolved quadruple system - where the dimmer component is in an edge-on configuration, could be producing transits detected by \textit{Kepler-K2} or \textit{TESS}, but the brighter pair is what is being picked up by Gaia DR3 NSS as a spectroscopic binary.

It's also possible that the reported radial velocity solution is an alias of the true period, complicating the situation further. 

With these caveats in mind, it's quite possible that one may not be able to associate a transit period with a Gaia DR3 NSS SB1 period, and this \textit{may not} be taken as the ability to rule out an eclipsing binary nature for the system in question.

As a result, we only claim to rule out planetary nature around a subset of matches between Gaia DR3 NSS SB1 systems and transiting planet candidates. In all other cases, we claim no further constraints on their nature.

\subsection{Single-Transit Candidates}
In situations where there is a Gaia SB1 solution for a star with a candidate single-transit planet, and knowing that there are mismatches between the SB1 period and the \textit{TESS} or \textit{Kepler} period for some systems, we cannot, therefore, determine that the transit event can be explained by an eclipse in the system \textit{as reported} by Gaia DR3. This is despite it appearing intuitively to be the case if the Gaia period is longer than the TESS observation baseline. From here the only clear path forward is to calculate the predicted transit times from the Gaia SB1 solution and compare these to the single TESS-reported transit time. For completeness, this exercise is also done for systems with known orbital periods, however in such cases it may be much less constraining.

For single-transit systems where we \textit{do} find a transit time match, we cannot neglect the possibility that the period given by the Gaia DR3 NSS SB1 solution is \textit{still} incorrect - perhaps an integer ratio alias of the true period. This doesn't change the results of the present work, however. We seek not to fully characterize the orbits of these eclipsing binaries, but simply to identify them among the list of transiting planet candidates.

The need to evaluate single-transit systems is a much more common problem for \textit{TESS} systems than for \textit{Kepler-K2} systems, owing to the latter's much longer observation baseline.

\subsection{Transit Time Predictions}
We must calculate the predicted transit times for Gaia DR3 NSS spectroscopic binaries to better evaluate whether or not they are responsible for the detection of a candidate planet in the \textit{TESS} data.

We first calculate the true anomaly as $f = \pi/2 - \omega$, where $\omega$ is the longitude of periapsis. The eccentric anomaly $E$ is then related to the eccentricity $e$ via
\[E = 2 \arctan \left( \sqrt\frac{1-e}{1+e}\,\tan{\frac{f}{2}}\right)\]

The time since periapsis passage is calculated from the orbital period $P$ via
\[T_e = \frac{P}{2 \pi} (E - e\sin{E})\]

The time of conjunction $T_{\text{conj}}$ is simply $T_{\text{Peri}} + T_e$. Whether or not this conjunction corresponds to a transit in an edge-on orbit is what TESS will determine, and is what we're especially interested in here.

We can also calculate the time of secondary eclipse by calculating the true anomaly as $f = 3 \pi/2 - \omega$. This might be relevant where only the secondary eclipse of a long-period eclipsing binary was detected in the TESS observations.

There are situations where a Gaia DR3 NSS SB1 system identifier corresponds to a transiting planet candidate system \textit{and} has the same orbital period, however the calculated conjunction time may be inconsistent with the transit time reported by \textit{TESS} or \textit{Kepler}. Given the potential issues with the Gaia DR3 NSS SB1 fitting, this appears to us to be indicative of possible poor fitting of the radial velocity. In situations where the eccentricity is low, the longitude of periapsis may be poorly constrained, which necessarily leads to high levels of uncertainty on the estimated transit time. The propagation of the error on the transit time from the Gaia SB1 solution may become such a large fraction of the period of the orbit that a match between it and the transit time reported by \textit{Kepler-K2} or \textit{TESS} may be non-constraining. Furthermore if the fitted period is an alias of the true period, then the transit window could easily not correspond to a true transit detection of the companion. 

As a result, while the conjunction epoch matching with the transit time is potentially a powerful indicator of the nature of the system, it is our least reliable test.

In some cases, the predicted transit time from the Gaia DR3 NSS SB1 solution falls $\sim$half an orbital period away from the transit time identified by \textit{TESS} or \textit{K2}. It's possible that this indicates that the detected transit is in fact a \textit{secondary} eclipse in a system that is inclined and eccentric such that no primary eclipses occur. Given the limited information we have to work with, we don't attempt to assess this in each case.

We find particularly poor matching for systems with low orbital periods from transit. This is likely due to the rather wide sampling of the system's radial velocity from Gaia, resulting in rather severe sample aliasing. As an illustration of this issue, the known eclipsing binary EPIC 202971774 \citep{Adams2016} is reported in Gaia DR3 NSS SB1 as having a radial velocity period of $658.2135\pm48.9873$ days,  however it's period from \textit{Kepler} is $\sim$0.2359 days. 

\section{Discussion of Specific Systems}

\subsection{EPIC 201257461}
The NASA Exoplanet Archive lists this system with a 50.278 day orbit, matching the Gaia SB1 solution. However, \cite{Armstrong2015} analyzed this system and reported a period of approximately half that value ($\sim$25.160 days). The given transit time is predicted transit time from the Gaia SB1 solution is $\sim$25-days off (and extrapolating the errors backward, the uncertainty on T$_\text{Transit}$ from the Gaia SB1 solution is only 3.7 percent of the 50-day orbital period, about 4.1 days). Because the stars appear to be a G+M binary, and because no primary or secondary eclipse would occur at the time of the Gaia $T_\text{conj}$ for a period of 50 days, we reject the 50-day period given at the NASA Exoplanet Archive in favor of that of \cite{Armstrong2015}. Substituting the Gaia SB1 period for the latter, we find that the Gaia prediction is a good match for the observed transit time.

\subsection{EPIC 211808055}
The NASA Exoplanet Archive lists this system as having a planet candidate in a 3.382 day orbit. Our Gaia DR3 NSS SB1 coordinate search returned a system at Gaia DR3 659288410630096000, with a period of $6.76224\pm0.00068$ days. This is exactly half (period ratio $0.5001\pm0.0001$) of the period of the planet candidate. However, Gaia DR3 659288410630096000 is actually EPIC 211807843, identified by \cite{Kruse2019} as an eclipsing binary in a 6.7642 day period. These are not the same star, yet the extraordinary coincidence of the orbital periods is difficult to ignore. The transit epochs given by the NASA Exoplanet Archive for EPIC 211808055 and by \cite{Kruse2019} for EPIC 211807843 also match. Therefore, we consider this system to be a blend, and rule out the planetary nature of EPIC 211808055's planet candidate.

\subsection{EPIC 211839430 and EPIC 211839462}
EPIC 211839430 and EPIC 211839462 correspond to BD+18 2050A and BD+18 2050B, respectively. They have the same 2.6128 day orbital period in the NASA Exoplanet Archive and essentially the same transit time. Neither of them correspond to the Gaia NSS SB1 source system returned by our coordinate search. Nonetheless, the secondary component was identified by \cite{Pepper2008} as an eclipsing binary with a 5.2301 day period. We consider both of these planet candidates to be clearly non-planetary in nature. The planet candidate for EPIC 211839430 doesn't exist, and is blended with the eclipsing binary EPIC 211839462 with twice the period. Unfortunately Gaia DR3 did not spectroscopically detect this eclipsing binary.

\subsection{EPIC 211972627}
This system is a known spectroscopic binary from \cite{Mermilliod2009} and listed in the Ninth Spectroscopic Binary Catalogue \citep{SB9} as SBC9 3451. Gaia DR3 recovers this spectroscopic binary and the NSS SB1 list for this star has parameters consistent with SB9. This system appears to be a 76.7 day binary G+M binary in the Praesepe. The NASA Exoplanet archive reports this system as having an $\sim$earth-sized planet in a $\sim$1.1 day orbit. This would be a rather interesting discovery if it could be confirmed. Notes in the Observation Logs for this system on the ExoFOP suggest it appears that the transit signal is likely to be electronic crosstalk contamination from the eclipsing binary S Cnc.

\subsection{EPIC 212024647 and EPIC 212024672}
EPIC 212024647 is a clear case of an eclipsing binary. It corresponds to Gaia DR3 663154156074506752, which Gaia DR3 NSS lists as a spectroscopic binary with a period ratio within $10^{-4}$ of unity. We notice that the nearby system EPIC 212024672 has again the same orbital period, as well as the same transit time reported by K2 for EPIC 212024647. Therefore, equipped with the new understanding of EPIC 212024647 as an eclipsing binary, we reject the planetary nature of both systems, with the latter being a blend with the former. 

\subsection{EPIC 220209578}
EPIC 220209578 is reported to have a candidate transiting planet in a $8.9045\pm0.0002$ day orbit. There is a brighter star to the North-West, Gaia DR3 2534988933319166976, which is identified by Gaia DR3 NSS has a spectroscopic binary in a $17.68055\pm0.02328$ day orbit (for a period ratio of $\sim0.5036$. This \textit{looks} like a blend with an eclipsing binary with twice the reported period, and it was identified as TOI-844 by \textit{TESS} at the $8.9$ day period. Eclipsing binaries often appear to have half their true orbital period in light curves as secondary eclipses can sometimes appear similar to primary eclipses. A complicating factor here, though, is that preliminary spectroscopic follow-up work performed by the ExoFOP's TFOPWG finds a centroid offset not on the Gaia-identified spectroscopic binary, but rather on EPIC 220209578 itself. The nature of this confusion is unclear -- are both systems eclipsing binaries with a coincidental 1:2 ratio periods? -- but the weight of the evidence strongly suggests that the transit-like events identified by Kepler-K2 on EPIC 220209578 are an astrophysical false positive and not a genuine planet.

\subsection{EPIC 228968232 (K2-256)}
One of the most interesting and important findings of this work centers around EPIC 228968232. The system was statistically validated by \cite{Livingston2018} as having a $2.63^{+0.25}_{-0.21}$ earth-radius planet in a $5.52011^{0.0024}_{0.0029}$ day orbit. This work earned the system a designation of K2-256. Our Gaia DR3 NSS SB1 coordinate search returned a candidate match on the star Gaia DR3 3693923014184650496 (TYC 4941-622-1). This star lies to the south-west at a distance of $\sim$78.51'' from K2-256 and is significantly brighter (V$\sim$10.46 compared to V$\sim$14.94). Gaia DR3 NSS SB1 lists this as a spectroscopic binary with a period of $5.1872\pm0.00018$ days, highly consistent with the orbital period of K2-256 b (period ratio $1.0003\pm0.0004$). Our epoch matching finds somewhat reasonable agreement as well. The transit mid-time reported at the NASA Exoplanet Archive is 2457586.5247, while our calculated transit mid-time from the Gaia DR3 SB1 solution that comes nearest this time is $2457585.47068\pm0.6908$. This is off by about a day, though consistent to within 2$\sigma$. 

K2-256 b's confirmation was arrived at by statistical validation. This eclipsing binary may have been stealthy enough to have evaded the even-odd test performed by \cite{Livingston2018}. Indeed other statistically validated exoplanets from Kepler-K2 have been determined to be blends with nearby eclipsing binaries. \cite{Shporer2017} determined that the statistically validated planets at K2-51, K2-67 and K2-76 were eclipsing binaries with low-mass stars, and \cite{Cabrera2017} showed that the statistically validated planets at K2-78, K2-82 and K2-92 were actually caused by nearby contaminating eclipsing binaries. The latter of these very much mirrors our proposed situation for K2-256. Consequently, we suggest that K2-256 b does not exist, and that this joins K2-78, K2-82 and K2-92 in being caused by an eclipsing binary blend.

Given the significance of such a claim, we wanted to verify that TYC 4941-622-1 is indeed an eclipsing binary with the reported period in Gaia DR3 NSS SB1. We viewed the Kepler-K2 photometry for the star in the MAST (Mikulski Archive for Space Telescopes) Portal and discovered a clear, classical eclipsing binary phase curve at the period of K2-256 b. While the primary eclipse was identified in the Kepler-K2 data for K2-256, the secondary eclipse appears to have been too faint in the blend scenario, allowing K2-256 to have survived an even-odd test. Altogether this gives us the confidence we need to argue that K2-256 b is not a real planet.

\begin{figure}
    \centering
    \includegraphics[width=85mm,scale=0.5]{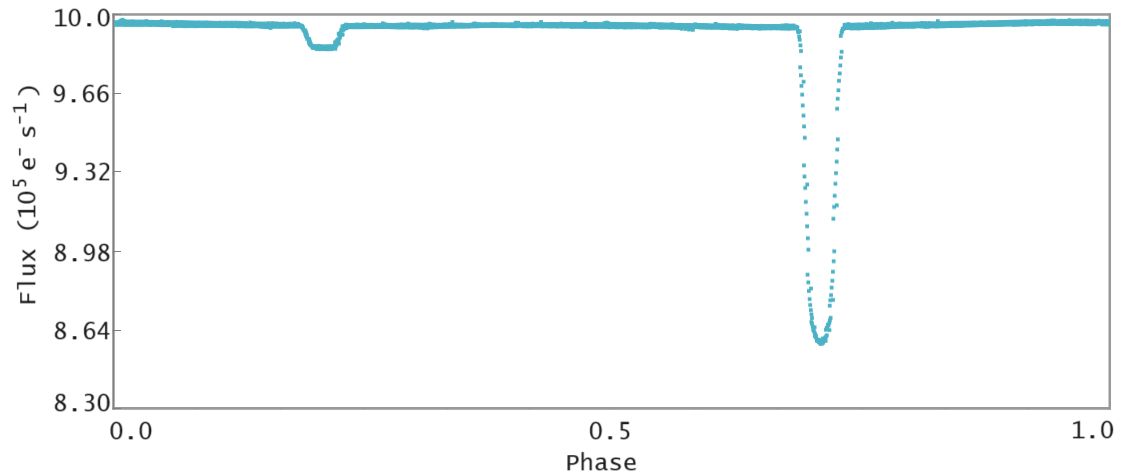}
    \caption{The \textit{Kepler-K2} Light Curve of the spectroscopic binary TYC 4941-622-1 phased to the period of the nearby ($\sim$78.51'') exoplanet system K2-256.}
    \label{fig:TYC 4941-622-1 Light Curve}
    \vspace{1.0cm}
\end{figure}

\subsection{TOI-289}
TOI-289 is a single-transit event detected by TESS. Gaia DR3 NSS SB1 lists the system as being a spectroscopic binary with a $224.706\pm0.906$ days. The predicted transit time (JD $2458367.0396\pm3.406$) does not match the observed transit time from TESS (JD $2458326.64586$), but the predicted secondary eclipse time (JD $2458318.7491\pm3.1553$) is a rather close match. The difference is $\sim7.897$ days, or about 4 percent of the orbital period. Interpreting the transit detection by TESS as the secondary eclipse of a long-period eclipsing binary explains the rather large transit depth, suggestive of a $\sim1.4\,R\text{J}$ object. These radii, while more consistent with the inflated hot Jupiter population, are not consistent with the radii for giant transiting planets at orbital periods sufficiently long to make TOI-289.01 a single-transit object.

TOI-289 was the subject of investigation by \cite{Holl2023} who provided updated orbital parameters based on the Gaia data. They provided a revised and more precise orbital period and eccentricity for this system. Using their orbital parameters and re-calculating the eclipse time, we find a new prediction of JD $2458320.7856\pm5.6759$, bringing this into consistency with the \textit{TESS} observed transit time. This confirms the system as an eclipsing binary. We report these more precise parameters in place of the default Gaia DR3 NSS SB1 parameters in this work.

\subsection{TOI-2473}
The ExoFOP website lists TOI-2473 as having a candidate transiting planet with a 1.2312-day orbit, as well as a second candidate with a 6.48369-day orbit. Our Gaia DR3 NSS SB1 coordinate matching brought our attention to the unrelated foreground star 2MASS J04573259+1025178, which is reported to be a spectroscopic binary in a $6.48369\pm0.00077$ day orbit. The similarity with the second planet candidate is hard to ignore and we suggest that this second candidate is the result of a nearby eclipsing binary blend. Unfortunately we're unable to achieve meaningful constraints from a match of the transit window from the Gaia orbit solution to the transit time observed by TESS because the uncertainty in the transit window is two-thirds the orbital period. We are unable to comment on the nature of the 1.2312-day planet candidate around TOI-2473.

\subsection{TOI-2482}
TOI-2482 was matched to a Gaia DR3 NSS SB1 entry of Gaia DR3 3173374317940622976. The \textit{Gaia} period is $7.6680\pm0.00326$ days, compared to the \textit{TESS} period of $14.9316\pm0.0002$ days. With a ratio of $1.9473\pm0.0008$, this is close to a factor of two, but \textit{not} close enough for us to consider it conclusively an eclipsing binary with either of the reported periods. The uncertainty on the transit window from the SB1 orbit solution is also much too large to attempt epoch-matching. The radial velocity semi-amplitude given by \textit{Gaia} of $K_\text{RV}=1.314\pm0.188\,\text{km s}^{-1}$, which at the \textit{TESS} period corresponds to a mass of $\sim13\,M_\text{J}$, and at the \textit{Gaia} period corresponds to a mass of $\sim10.5\,M_\text{J}$. This possible low (perhaps planetary) mass has prompted us to comment on this system.

The given radius from \textit{TESS} ($\sim1.5\,R_{\oplus}$) is \textit{much} too small for either mass. This could be caused by a grazing transit in a high-impact parameter orbit or contamination from another star. The \textit{TESS}-reported transit duration of $\sim4.729$ hours also appears to make the shorter Gaia period difficult to explain for this late-K/early-M dwarf star. An inspection of the TESS light curve data shows one transit each in two TESS Sectors: Sector 5 and Sector 32. It's true period may be one of any number of candidate periods (half, 1/4th, etc the time between the two transits) and the Gaia period \textit{must} be wrong.

We cannot, therefore, determine the nature of this system, and we leave it as a tempting target for future work given it's (relatively) low $K_\text{RV}$.

\subsection{TOI-4330}
TOI-4330 is reported by \textit{TESS} as a $17 R_\text{J}$ planet candidate in a 3.3513 day orbit. The star corresponds to the star HD 129051, a K-type giant reported to have a stellar companion by \cite{Mugrauer2023}. Neither the TICv8.2 radius of $8.7 R_\odot$ nor the Gaia DR3 radius of $8.3 R_\odot$ appear to support the claim of a transiting planet in such a short period.

Our coordinate search returned a result at TYC 7304-483-1, which Gaia DR3 NSS SB1 lists as having a $K=54.1\pm2.6$ km s$^{-1}$ signal in a $3.36662\pm0.00034$ day orbit. This is close, but not quite equal to, the TESS period (period ratio of $0.9955\pm0.0001$). We examined the \textit{TESS} light curve for TYC 7304-483-1 and found eclipses at $\sim3.6$ days, but given the highly blended nature of the system, we remain unable to comment on its nature.

\subsection{TOI-951 and TOI-953}
TOI-951 and TOI-953 are the B and A components, respectively, of the visual binary system ADS 3266. Our Gaia DR3 NSS SB1 search produced the latter of these two as a match with TOI-953. The \textit{Gaia} source is reported as a spectroscopic binary with a period equal to that of the \textit{TESS} photometric period. The eclipsing binary nature of the star was also noted by \cite{Schanche2019}.

The same \textit{Gaia} source was also returned in our coordinate search for TOI-951, which \textit{also} has the same periods as TOI-953. TOI-951, therefore, appears to be a blend with the TOI-953 eclipsing binary. Cosnequently, we reject the planetary nature of both candidate systems.

\section{Conclusions}

Searching for Gaia DR3 NSS SB1 systems in the vicinity of known transiting planet candidates, we find numerous examples of situations where the candidate transiting planet can be established to be an eclipsing binary star, as well as some cases where a nearby eclipsing binary with the same orbital period can be determined to be blended with the candidate planetary system, in both events leading to a "false positive." 

Of particular note, we identify TYC 4941-622-1 as an eclipsing binary from the Gaia DR3 NSS combined with \textit{Kepler} light curve data, and note that it has the same period and ephemeris to the nearby ($\sim$78.51'') statistically validated transiting planet K2-256 b. This makes it almost certainly the case that K2-256 b is not a genuine planet, and reinforces an already-noted concerns about eclipsing binaries surviving statistical validation techniques. The flux of TYC 4941-622-1 affected the \textit{Kepler} data for K2-256 only slightly, such that only the primary eclipses were detected in the K2-256 b light curve. As a result, K2-256 b passed the even-odd test that would have rejected most eclipsing binary systems.

We find a higher eclipsing binary identification rate with \textit{TESS} planet candidates than we do for \textit{Kepler-K2} planet candidates. This is likely because the \textit{TESS} target systems are on average much brighter, and consequently more easily detected by Gaia in the event that they are indeed eclipsing binaries.

Numerous systems in Gaia DR3 NSS SB1 correspond to known transiting planet candidates but the spectroscopic solution from \textit{Gaia} does not match the photometric period from \textit{TESS}. While the \textit{Gaia} period is probably wrong, we suggest that these systems are nevertheless likely to be eclipsing binaries -- something massive \textit{must} be producing the radial velocity variability identified by \textit{Gaia}. 

Overall we find that Gaia DR3 NSS is a powerful tool for constraining the true nature of transiting planet candidates. While \textit{Gaia}'s radial velocity precision is essentially inadequate to the task of detecting planetary mass objects, it \textit{can} detect eclipsing binaries. A planet candidate system appearing in the Gaia DR3 NSS SB1 list should be considered an immediate warning, but SB1 stars in the surrounding area should also be investigated to help identify eclipsing binary blends that can occur at a variety of distances from the candidate planetary system.

\section{Acknowledgements}
This research has made use of the Exoplanet Follow-up Observation Program (ExoFOP; DOI: 10.26134/ExoFOP5) website, which is operated by the California Institute of Technology, under contract with the National Aeronautics and Space Administration under the Exoplanet Exploration Program.

\LongTables
\clearpage
\begin{deluxetable*}{lrcccrr}
\tablecolumns{7}
\tablecaption{Gaia DR3 NSS SB1 Coordinate Matches with Kepler-K2 Candidate Systems}
\tablehead{\colhead{EPIC ID} & \colhead{Gaia DR3 ID} & \colhead{Identifier?}& \colhead{Period?} & \colhead{Epoch?} & \colhead{Nature} & \colhead{Note}}
\startdata
EPIC 201160662 & 3599805123787755648 & Yes & Yes & ? & EB & \cite{Kruse2019}\\
EPIC 201257461 & 3601830428502237312 & Yes & Yes & Yes & EB & \cite{Armstrong2015}\\
EPIC 201516974 & 3798898062211776256 & Yes & Yes & No & EB & New EB\\
EPIC 202088212 & 3368700905049734784 & Yes & No & No & ? \\
EPIC 202971774 & 6045243267437785856 & Yes & No & No & EB & \cite{Adams2016}\\
EPIC 203485624 & 6048670406520239104 & Yes & Yes & ? & EB & \cite{Drake2017}\\
EPIC 203942067 & 6049731027225353344 & Yes & No & ? & EB &\cite{Kruse2019}\\
EPIC 204546592 & 6243506654403685504 & Yes & No & ? & ? & \\
EPIC 211002562 & 61904742323780992 & Yes & No & ? & ? & \\
EPIC 211093684 & 66747816167123712 & Yes & Yes & ? & EB & V1283 Tau\\
EPIC 211342524 & 600099122725424768 &Yes&Yes&Yes&EB&New EB\\
EPIC 211408138 & 604911410242410752 &Yes&Yes&Yes&EB&HV Cnc\\
EPIC 211418290 & 601790961883658112 &Yes&Yes&Yes?&EB&New EB\\
EPIC 211424769 & 602552128870918144 &Yes&Yes&Yes&EB&New EB\\
EPIC 211452175 & 604460473035451904 &Yes&Yes&Yes&EB&New EB\\
EPIC 211642307 & 655332195994611584 &Yes&No&No&?&\\
EPIC 211705654 & 610081416995398656 &Yes&No&No&?&\\
EPIC 211710534 & 610074785565868032 &Yes&Yes&?&EB&\cite{Kruse2019}\\
EPIC 211808055 & 659288410630096000 &No&Yes&Yes?&Blend&See discussion\\
EPIC 211830293 & 611323517241144576 &Yes&Yes&Yes&EB&New EB\\
EPIC 211839430 & 659847065615319936 &No&No&No&EB&\cite{Pepper2008}\\
EPIC 211839462 & 659847065615319936 &No&No&No&EB&\cite{Pepper2008}\\
EPIC 211886472 & 635409251456055168 &Yes&No&No&EB&New EB\\
EPIC 211972627 & 661302887731820416 &Yes&No&No&Blend?&See discussion\\
EPIC 212024647 & 663154156074506752 &Yes&Yes&?&EB&New EB\\
EPIC 212024672 & 663154156074506752 &No&Yes&?&Blend&See discussion\\
EPIC 212033577 & 637075282154501760 &Yes&Yes&Yes&EB&New EB\\
EPIC 212039539 & 665412587317415296 &Yes&Yes&?&EB&New EB\\
EPIC 212069706 & 685756434352272896 &No&No&?&EB&\cite{Adams2016}\\
EPIC 212117087 & 665841018894658688 &Yes&Yes&?&EB&\cite{Yu_2018}\\
EPIC 212772313 & 3631703609673030784&Yes&No&No&?\\
EPIC 214912104 & 6769386779452266240&No&Yes&Yes?&Blend\\
EPIC 220209578 & 2534988933319166976&No&Yes&?&Blend?&See discussion\\
EPIC 220477223 & 2576503365647719296&Yes&Yes&Yes&EB&\cite{Kruse2019}\\
EPIC 228729473 & 3579529686991039360&Yes&Yes&Yes?&EB&New EB\\
EPIC 228968232 & 3693923014184650496&No&Yes&Yes&Blend&K2-256 - See discussion\\
EPIC 247281516 & 3408649117423934208&Yes&Yes&Yes?&EB&New EB\\
EPIC 247724061 & 146833837950817408&Yes&No&No&?&\\
EPIC 248740905 & 3876036185948161792&Yes&Yes&Yes&EB&New EB\\
EPIC 249239123 & 6238897432936164736&Yes&Yes&?&EB&New EB\\
EPIC 251292838 & 630850328354376320&Yes&Yes&No?&EB&New EB\\
EPIC 251330444 & 635898431051113216 &Yes&Yes&?&EB&\cite{Yu_2018}\\
EPIC 251380988 & 687194870439222656&Yes&No&No&?\\
\vspace{-0.3cm}
\enddata
\tablecomments{Matches between Gaia DR3 NSS SB1 and Kepler K2 candidate systems. Systems identified by this work are marked as 'New EB.' Systems where there was no apparent association, either ID, period or epoch, are clearly field SB1s and have not been included here.}
\end{deluxetable*}
\clearpage
\clearpage
\begin{deluxetable*}{lrcccrr}
\tablecolumns{7}
\tablecaption{Gaia DR3 NSS SB1 Coordinate Matches with TESS Object of Interest (TOI) Systems}
\tablehead{\colhead{TESS ID} & \colhead{Matched Gaia DR3 ID} & \colhead{Identifier?}& \colhead{Period?} & \colhead{Epoch?} & \colhead{Nature} & \colhead{Note}}
\startdata
TOI-121 & 6571502449115103232 &Yes&Yes&Yes&EB&\cite{Triaud2017}\\
TOI-148 & 6594248591617970560 &Yes&Yes&?&EB&\cite{Grieves2021}\\
TOI-156 & 4720986721294105216 &Yes&Yes&Yes&EB&New EB\\
TOI-158 & 6508105536369837312 &Yes&No&?&?\\
TOI-162 & 6804708139515008640 &Yes&Yes&?&EB&New EB\\
TOI-184 & 5270358663889956480 &Yes&Yes&?&EB&New EB\\
TOI-185 & 4955371367334610048 &Yes&Yes&?&Pl&\cite{Hellier2009}\\
TOI-222 & 6531037981670835584 &Yes&?&No&?\\
TOI-236 & 2434381179836445952 &Yes&Yes&?&EB&\cite{Triaud2017}\\
TOI-276 & 4769981165586553728 &Yes&Yes&?&EB&\cite{Triaud2017}\\
TOI-289 & 4919562197762515456 &Yes&?&No&EB&See discussion\\
TOI-383 & 5280337144228825472 &No&No&?&Blend&\cite{Grieves2021}\\
TOI-416 & 3262082160152897408 &Yes&Yes&No?&EB&New EB\\
TOI-425 & 2988427421239442432 &Yes&Yes&?&EB&New EB\\
TOI-446 & 4824755364549309824 &Yes&Yes&?&EB&\cite{Psaridi2022}\\
TOI-459 & 4676885550423032064 &Yes&Yes&Yes&EB&New EB\\
TOI-478 & 2919669912574582912 &Yes&Yes&?&EB&\cite{Psaridi2022}\\
TOI-484 & 3220802537980446848 &No&Yes&?&Blend&New EB\\
TOI-503 & 650254479499119232 &Yes&Yes&?&EB&\cite{Subjak2020}\\
TOI-586 & 5538255324833781888 &Yes&Yes&Yes&EB&New EB\\
TOI-592 & 5695996352497664512 &Yes&No&?&?\\
TOI-604 & 5596318442261247232 &Yes&No&?&?\\
TOI-646 & 4641589646622686720 &Yes&Yes&?&EB&New EB\\
TOI-665 & 5417424326998593408 &Yes&Yes&Yes&EB&New EB\\
TOI-668 & 5416752486737577984 &Yes&Yes&?&EB&New EB\\
TOI-691 & 5569479810090831104 &Yes&Yes&?&?\\
TOI-707 & 4779298186602189440 &Yes&Yes&No?&EB&EBLM J0418-53\\
TOI-746 & 5280337144228825472 &Yes&Yes&?&EB&\cite{Grieves2021}\\
TOI-759 & 5395959695358663296 &Yes&Yes&?&EB&New EB\\
TOI-764 & 5385762893242980992 &Yes&Yes&?&EB&\cite{Psaridi2022}\\
TOI-768 & 6194034953338162816 &Yes&No&?&?\\
TOI-827 & 5210891547437609856 &Yes&No&?&?\\
TOI-838 & 6161800502228330240 &Yes&Yes&Yes&EB&New EB\\
TOI-844 & 2534988933319166976 &Yes&No&?&?\\
TOI-846 & 4916461712411594112 &Yes&Yes&Yes&EB&New EB\\
TOI-901 & 5925532800797072128 &Yes&Yes&?&EB&New EB\\
TOI-924 & 4769649280577737728 &Yes&Yes&?&EB&New EB\\
TOI-948 & 2961066353557939456 &Yes&Yes&?&EB&New EB\\
TOI-951 & 3285409673726608768 &No&Yes&?&Blend&See discussion\\
TOI-953 & 3285409673726608768 &Yes&Yes&?&EB&See discussion\\
TOI-996 & 5714269239318650880 &Yes&Yes&?&EB&New EB\\
TOI-1018 & 5505717481529404416 &Yes&Yes&?&EB&New EB\\
TOI-1035 & 251077357420917376 &Yes&No&No&?\\
TOI-1059 & 5911682531996474624 &Yes&Yes&Yes&EB&New EB\\
TOI-1072 & 4649937757389347456 &Yes&Yes&Yes&EB&New EB\\
TOI-1111 & 6445176465824182272 &Yes&Yes&Yes?&EB&New EB\\
TOI-1115 & 6650192366014887424 &Yes&Yes&?&EB&New EB\\
TOI-1119 & 6660655834058823680 &Yes&Yes&Yes&EB&New EB\\
TOI-1140 & 1421127098253596288 &Yes&Yes&Yes?&EB&New EB\\
TOI-1141 & 1710280239538684928 &Yes&Yes&Yes&EB&New EB\\
TOI-1153 & 1726947014749701504 &Yes&Yes&No?&EB&New EB\\
TOI-1186 & 1635325672880724864 &Yes&Yes&Yes&EB&New EB\\
TOI-1251 & 2160710158704303360 &Yes&Yes&?&EB&New EB\\
TOI-1270 & 2289068176923197952 &Yes&Yes&No&EB&New EB\\
TOI-1338 & 5494443978353833088 &Yes&No&No&EB&\cite{Triaud2017}\\
TOI-1341 & 2236532858514382336 &Yes&Yes&Yes&EB&\cite{Schanche2019}\\
TOI-1371 & 1898569131896247680 &Yes&Yes&Yes?&EB&\cite{Schanche2019}\\
TOI-1393 & 2004338572785772800 &Yes&No&No&?\\
TOI-1455 & 2246355002042750976 &Yes&Yes&?&EB&New EB\\
TOI-1461 & 320523552351605632 &Yes&Yes&?&EB&\cite{Schanche2019}\\
TOI-1514 & 2209081592227687680 &Yes&No&?&?\\
TOI-1553 & 348551718652896000 &Yes&Yes&Yes&EB&New EB\\
TOI-1567 & 443040655569966208 &Yes&?&Yes&EB&New EB\\
TOI-1572 & 352678495029040000 &Yes&2x&Yes&EB&New EB\\
TOI-1591 & 462926457230782208 &Yes&?&Yes&EB&New EB\\
TOI-1608 & 125548289270234880 &Yes&Yes&?&EB&V687 Per\\
TOI-1645 & 345093337969643392 &Yes&Yes&?&EB&\cite{Schanche2019}\\
TOI-1656 & 275228582433787904 &Yes&?&No&?\\
TOI-1677 & 473104258412065408 &Yes&No&No&?\\
TOI-1698 & 957955110633719552 &Yes&2x&Yes&EB&New EB\\
TOI-1711 & 932236090550183296 &Yes&2x&Yes&EB&\cite{Maxted2023}\\
TOI-1712 & 917823348536224768 &Yes&Yes&?&EB&\cite{Schanche2019}\\
TOI-1786 & 1039196547341014016 &Yes&Yes&?&EB&New EB\\
TOI-1787 & 802616348383896448 &Yes&Yes&?&EB&New EB\\
TOI-1788 & 751544445585226496 &Yes&Yes&Yes&EB&New EB\\
TOI-1822 & 764958758026610560 &Yes&Yes&Yes?&EB&New EB\\
TOI-1834 & 1468504126582376448 &Yes&No&?&?\\
TOI-1856 & 1602005522755589760 &Yes&2x&No&EB&New EB\\
TOI-1888 & 5254794943714396800 &No&Yes&Yes?&Blend&New EB\\
TOI-1897 & 548075249022607744 &Yes&Yes&No&EB&New EB\\
TOI-1951 & 5199893510584259968 &Yes&Yes&Yes&EB&New EB\\
TOI-1952 & 5199213874958918784 &Yes&No&?&?\\
TOI-1974 & 5235114127904811648 &Yes&Yes&?&EB&New EB\\
TOI-1991 & 5348534425968598400 &Yes&Yes&?&EB&New EB\\
TOI-1996 & 5377428178491618944 &Yes&2x&?&EB&New EB\\
TOI-2008 & 5096613016130459136 &Yes&?&Yes&EB&New EB\\
TOI-2028 & 2204963130906718592 &Yes&No&No&?\\
TOI-2038 & 2295415623189777024 &Yes&Yes&?&EB&New EB\\
TOI-2065 & 3961485079994112896 &Yes&?&No&EB&\cite{Collins2018}\\
TOI-2121 & 1342147666203453312 &Yes&No&?&?\\
TOI-2157 & 4570820611494681984 &Yes&Yes&Yes&EB&New EB\\
TOI-2159 & 1363495509089066880 &Yes&Yes&Yes&EB&\cite{Schanche2019}\\
TOI-2218 & 5285592504869495296 &Yes&No&?&?\\
TOI-2335 & 4852751438852929920 &Yes&No&?&?\\
TOI-2349 & 5689821044216119296 &Yes&Yes&Yes?&EB&New EB\\
TOI-2360 & 4671762272913404160 &Yes&Yes&No&EB&New EB\\
TOI-2366 & 3023791185522078592 &Yes&Yes&Yes&EB&New EB\\
TOI-2413 & 6825245573613327616 &Yes&Yes&Yes&EB&New EB\\
TOI-2442 & 4884422424614580352 &Yes&Yes&Yes&EB&New EB\\
TOI-2469 & 3173208463482530176 &Yes&Yes&?&EB&New EB\\
TOI-2473 & 3291551648758101248 &No&Yes&?&Blend&See discussion\\
TOI-2482 & 3173374317940622976 &Yes&2x?&?&?&See discussion\\
TOI-2489 & 6567569977058424448 &Yes&No&?&?\\
TOI-2490 & 4818818585874648320 &Yes&No&?&?\\
TOI-2492 & 3321419774751152128 &Yes&Yes&?&EB&New EB\\
TOI-2521 & 3005862518856922752 &Yes&No&?&?\\
TOI-2533 & 1259922196651254272 &Yes&Yes&?&EB&New EB\\
TOI-2543 & 579084083968560256 &Yes&No&?&EB&\cite{Psaridi2022}\\
TOI-2544 & 5320246568758675328 &Yes&2x&Yes&EB&New EB\\
TOI-2546 & 3147890181029001984 &Yes&No&?&?\\
TOI-2555 & 5697796768432105216 &Yes&Yes&Yes&EB&New EB\\
TOI-2598 & 3026047834355902464 &Yes&Yes&Yes&EB&New EB\\
TOI-2606 & 5301643549272386560 &Yes&Yes&?&EB&New EB\\
TOI-2611 & 6404469659144989056 &Yes&No&?&?\\
TOI-2614 & 4635190862611016448 &Yes&No&?&?\\
TOI-2664 & 3855170410190523008 &Yes&Yes&?&EB&New EB\\
TOI-2674 & 3544898334892093952 &Yes&No&?&?\\
TOI-2819 & 3166138805578006016 &Yes&Yes&?&EB&New EB\\
TOI-3006 & 5310692633056834432 &Yes&Yes&Yes&EB&New EB\\
TOI-3010 & 5299710470390857728 &Yes&Yes&Yes?&EB&New EB\\
TOI-3148 & 5202007248674569728 &Yes&Yes&?&EB&New EB\\
TOI-3155 & 5869945139392446336 &Yes&Yes&?&EB&New EB\\
TOI-3191 & 5851691833256275200 &Yes&Yes&?&EB&New EB\\
TOI-3241 & 6070306974290936960 &Yes&Yes&?&EB&New EB\\
TOI-3243 & 5872390109623076992 &Yes&No&?&?\\
TOI-3260 & 4639398560466507904 &Yes&Yes&Yes?&EB&New EB\\
TOI-3298 & 6419210296144773504 &Yes&Yes&?&EB&New EB\\
TOI-3501 & 3486043573401367168 &Yes&Yes&?&EB&New EB\\
TOI-3502 & 5343051878733534592 &Yes&Yes&?&EB&New EB\\
TOI-3542 & 2057752233480414848 &Yes&Yes&?&EB&New EB\\
TOI-3697 & 409061893681183232 &Yes&No&?&?\\
TOI-3756 & 189461074831319680 &Yes&Yes&?&EB&New EB\\
TOI-3796 & 1087913090069066496 &Yes&Yes&?&EB&New EB\\
TOI-3904 & 1512692949146107520 &Yes&Yes&No?&EB&New EB\\
TOI-3995 & 420954215752408192 &Yes&Yes&?&EB&New EB\\
TOI-4226 & 5614312912124661632 &Yes&No&?&?\\
TOI-4245 & 5516850925816335744 &Yes&No&?&?\\
TOI-4330 & 6215224775792959232 &No&No&?&?&See discussion\\
TOI-4331 & 6005773858021162624 &Yes&No&?&?\\
TOI-4338 & 6063252821253530752 &Yes&Yes&Yes&EB&New EB\\
TOI-4400 & 6463801879263478144 &Yes&Yes&?&EB&New EB\\
TOI-4420 & 5808849641863958528 &No&Yes&?&EB&Blend\\
TOI-4462 & 4605954852723545088 &Yes&Yes&?&EB&New EB\\
TOI-4657 & 2395520590619923328 &Yes&Yes&?&EB&New EB\\
TOI-4771 & 3087689411148553472 &Yes&No&?&?\\
TOI-4792 & 577614689757449728 &Yes&Yes&Yes?&EB&New EB\\
TOI-4807 & 5587000905841875456 &Yes&Yes&?&EB&New EB\\
TOI-4973 & 5230545317792027776 &Yes&Yes&No&EB&New EB\\
TOI-5103 & 600788374781198720 &Yes&2x&Yes?&EB&New EB\\
TOI-5149 & 2167586710592820096 &Yes&Yes&Yes&EB&New EB\\
TOI-5156 & 3278323493085459456 &Yes&Yes&No&EB&New EB\\
TOI-5197 & 1130631384551617024 &Yes&Yes&?&EB&New EB\\
TOI-5200 & 2148571309814936576 &Yes&Yes&No&EB&New EB\\
TOI-5307 & 2581395161599229952 &Yes&Yes&Yes&EB&New EB\\
TOI-5372 & 109773424146717568 &Yes&Yes&?&EB&New EB\\
TOI-5379 & 983011056484954496 &Yes&Yes&?&EB&New EB\\
TOI-5394 & 623779716969345920 &Yes&Yes&Yes&EB&New EB\\
TOI-5427 & 3341401062125577088 &Yes&Yes&?&EB&\cite{Schmidt2023}\\
TOI-5475 & 3429797609407986176 &Yes&Yes&Yes?&EB&New EB\\
TOI-5547 & 674670853180191360 &Yes&Yes&Yes?&EB&New EB\\
TOI-5677 & 3643353863081886336 &Yes&Yes&?&EB&New EB\\
TOI-5746 & 1018729069670619520 &Yes&?&Yes&EB&New EB\\
TOI-5761 & 548058859427458688 &Yes&Yes&No?&EB&New EB\\
TOI-5811 & 1814257343328666624 &Yes&No&?&?\\
TOI-5860 & 1802021977935667200 &Yes&Yes&?&EB&New EB\\
TOI-5882 & 1869489729418662528 &Yes&Yes&?&EB&New EB\\
TOI-6066 & 1117014688955305984 &Yes&No&?&?\\
TOI-6096 & 3845758148275527936 &Yes&No&?&?\\
TOI-6106 & 5696528202583814400 &Yes&No&?&?\\
TOI-6111 & 2070280099026014336 &Yes&Yes&Yes&EB&New EB\\
TOI-6117 & 2838654459861938944 &Yes&Yes&?&EB&New EB\\
TOI-6133 & 1928695681777800960 &Yes&Yes&?&EB&New EB\\
TOI-6151 & 2167122162624970624 &Yes&No&?&?\\
TOI-6207 & 2871631592420978432 &Yes&Yes&?&EB&New EB\\
TOI-6210 & 1980382761343741952 &Yes&No&?&?\\
TOI-6218 & 392652614647892736 &Yes&Yes&?&EB&New EB\\
TOI-6242 & 2859788722613818112 &Yes&Yes&?&EN&New EB\\
TOI-6549 & 4768885261731108480 &Yes&No&?&?\\
TOI-6566 & 6197199790185469184 &Yes&No&?&?\\
TOI-6568 & 6057181524220247680 &Yes&Yes&Yes?&EB&New EB\\
TOI-6611 & 5795018365391402368 &Yes&Yes&?&EB&New EB\\
\vspace{-0.3cm}
\enddata
\tablecomments{Matches between Gaia DR3 NSS SB1 and TESS candidate systems. Systems identified by this work are marked as 'New EB'}
\end{deluxetable*}
\clearpage

\clearpage
\includegraphics[width=2.1\linewidth, trim={0cm 0cm 0 0cm}]{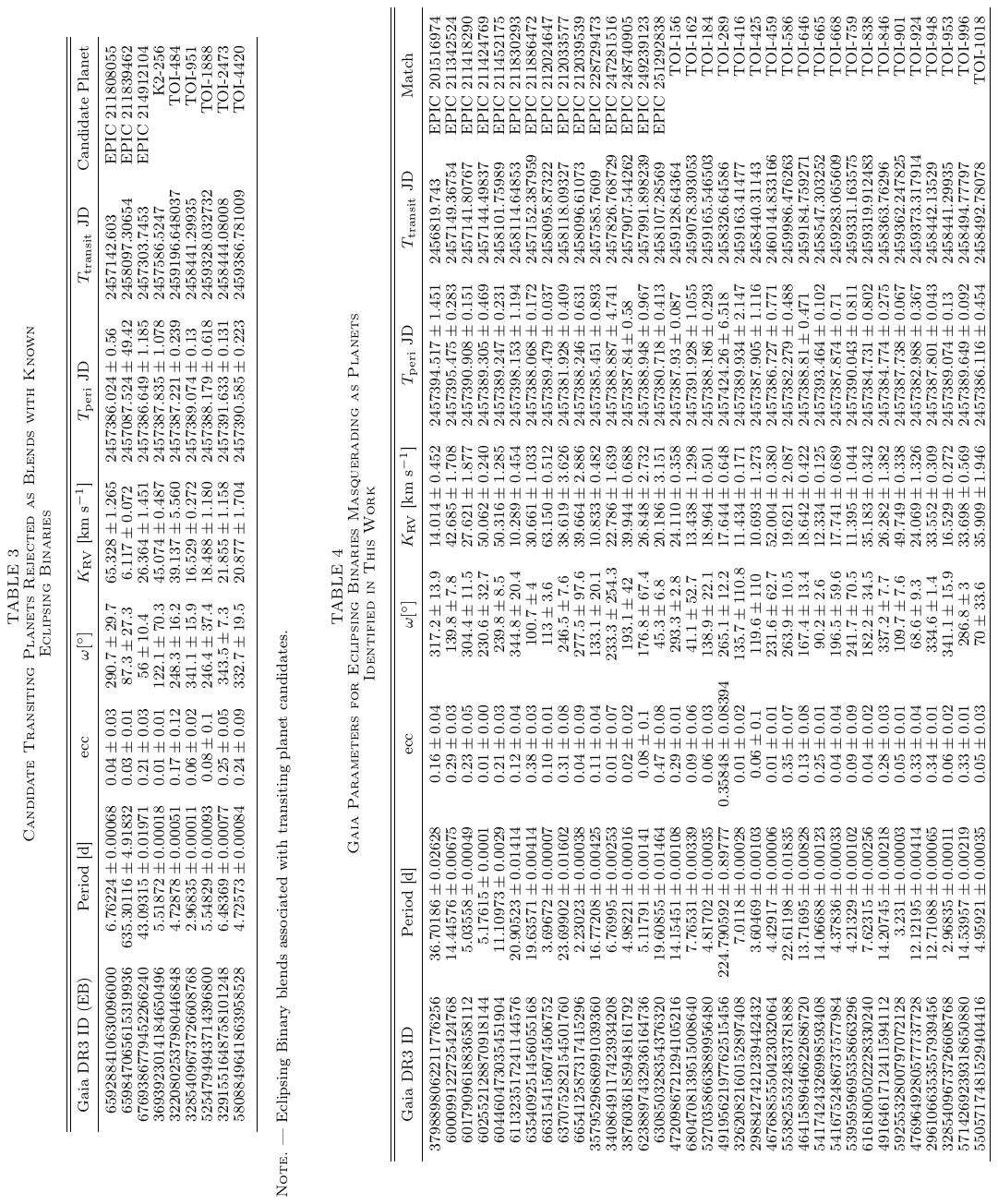}
\clearpage
\includegraphics[width=2.1\linewidth, trim={0cm 0cm 0 0cm}]{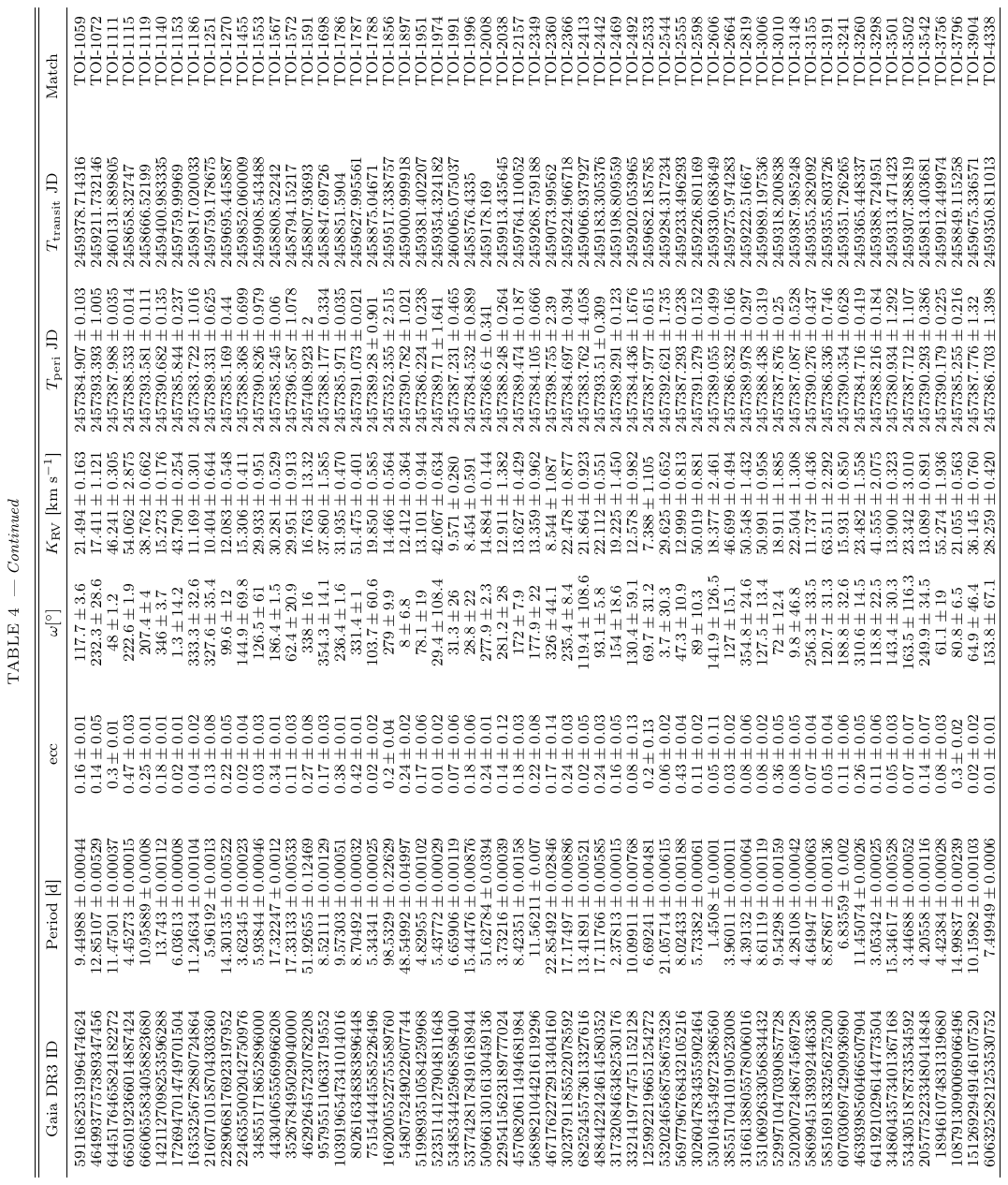}
\clearpage
\includegraphics[width=2.1\linewidth, trim={0cm 0cm 0 0cm}]{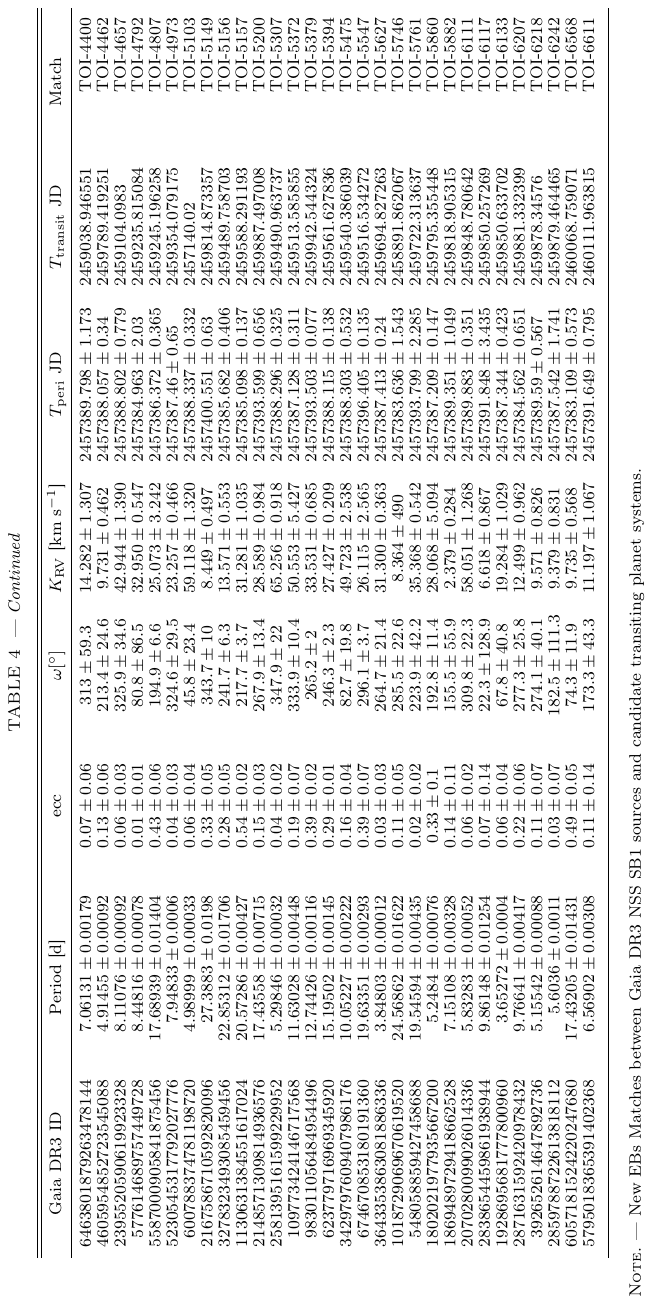}
\clearpage

\acknowledgements


\bibliography{main.bib}
\bibliographystyle{plainnat}

\clearpage

\end{document}